\documentclass[a4paper,fleqn]{cas-dc}

\usepackage[numbers]{natbib}
\usepackage{algorithm}
\usepackage{algorithmic}
\usepackage{mathtools}

\newdefinition{definition}{Definition}

\begin{document}
\let\WriteBookmarks\relax
\def\floatpagepagefraction{1}
\def\textpagefraction{.001}

\shorttitle{PromptShift: Identity-Cue Preference Drift in LLM Recommenders}

\shortauthors{Z. Gan and Q. Dong}

\title [mode = title]{Measuring and Mitigating Identity-Cue Preference Drift in LLM-based Recommender Systems}

\author[1]{Zhuoxiong Gan}
\credit{Conceptualization, Methodology, Software, Investigation, Writing -- original draft}

\affiliation[1]{organization={Institute of Intelligent Computing, University of Electronic Science and Technology of China},
            city={Chengdu},
            country={China}}

\author[1]{Qiang Dong}
\cormark[1]
\ead{dongq@uestc.edu.cn}
\credit{Supervision, Writing -- review \& editing, Funding acquisition}

\cortext[1]{Corresponding author}

\begin{abstract}
Large language model (LLM)-based recommender systems condition their outputs on users' natural-language interaction histories; however, identity cues embedded in prompts can steer recommendations toward group-level patterns even when the underlying behavioral evidence remains unchanged. We introduce PromptShift, an interpretable, training-free framework for quantifying and mitigating such identity-cue preference drift. We define Drift as the divergence---in both item membership and ranking order---between a recommendation list generated under an identity-cued prompt and the reference list produced from the same user's interaction history alone. An identity slice denotes the subset of users sharing a given value of a cued attribute. SliceShift then measures the extent to which a cued list gravitates, relative to the history-only reference, toward items that are more popular within the cued slice than among the global user population. To evaluate recommendation utility beyond conventional accuracy, we propose Difficulty@20, a difficulty-weighted hit metric that credits only relevant items, assigning higher credit to hits that are less popular within the cued slice and ranked higher in the list. All components are supported by an identity-slice-by-item table constructed from positive interactions, which further enables an adaptive post-hoc reranking strategy: the reranker interpolates between the original LLM ranking and inverse slice-popularity, with the interpolation weight personalized by each user's within-slice history mainstreamness---the average slice popularity of items in their interaction history. Experiments on MovieLens~1M and Last.fm~1K with three LLMs show that identity-cued prompts incur higher mean Drift than identity-free paraphrase controls---an effect beyond generic wording sensitivity---and that SliceShift is positive across all six dataset--model settings. PromptShift consistently reduces both Drift and SliceShift---lowering macro-mean SliceShift by 62.42\%---while improving Difficulty@20, HitRate@20, and MRR@20, at the cost of a slight decrease in NDCG@20. These results demonstrate that identity-cue preference drift can be measured and mitigated without any model training, albeit with a modest, metric-dependent utility cost.
\end{abstract}


\maketitle

\section{Introduction}
\label{sec:intro}

\begin{figure*}[t]
  \centering
  \includegraphics[width=\textwidth]{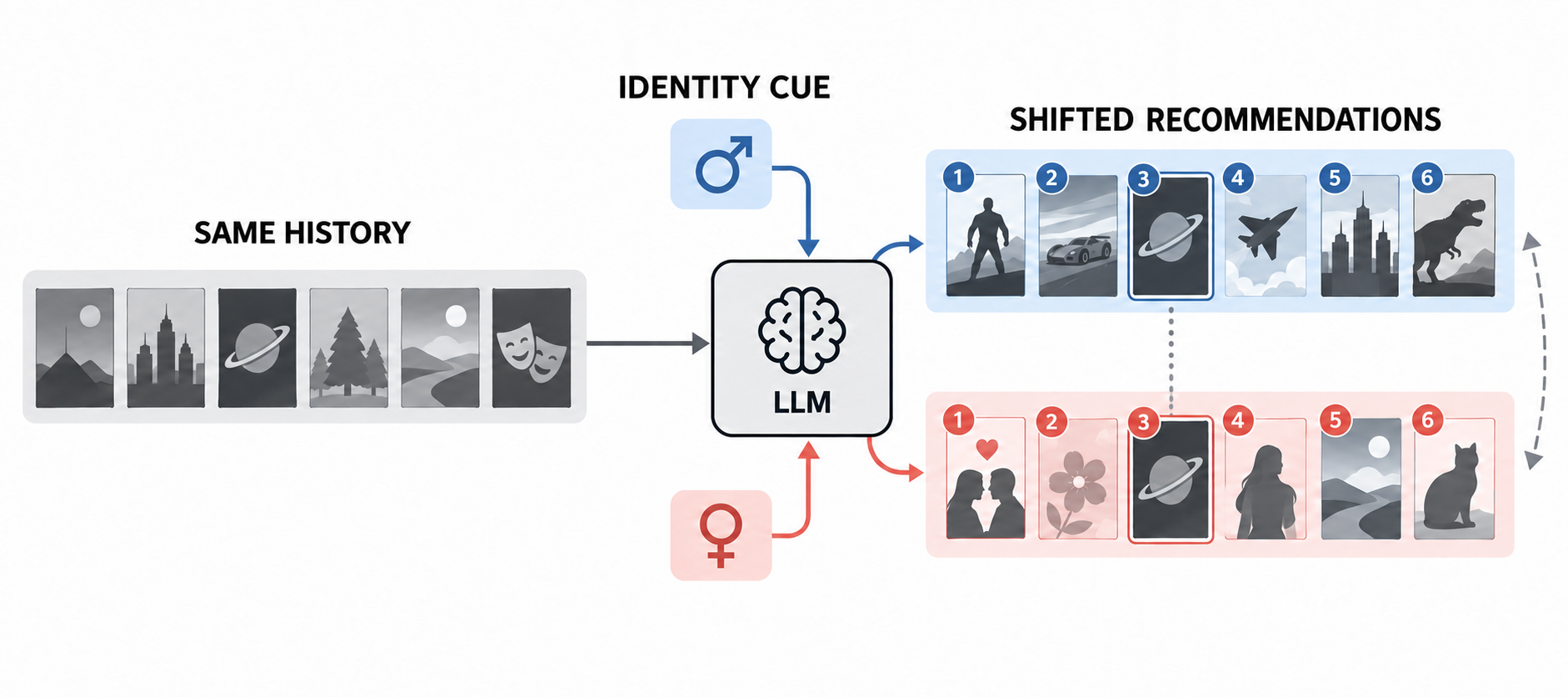}
  \caption{An illustration of identity-cue preference drift. Given the same interaction history, appending a single factual identity cue to the prompt (e.g., gender) shifts the LLM's recommendations toward items typical of the cued group (top: male cue; bottom: female cue), while items grounded in the user's history (the highlighted planet item) retain their positions in both lists.}
  \label{fig:teaser}
\end{figure*}

Recommender systems are increasingly built on large language models (LLMs), which take a user's interaction history---expressed in natural language---as input and directly generate a ranked list of items, without task-specific training~\cite{hou2024large,gao2023chatrec,dai2023uncovering,sanner2023large,lin2023how}. In practical deployments, however, the prompts issued to these models seldom consist of behavioral evidence alone. User profiles assembled from account records or persona descriptions routinely embed demographic attributes such as gender and age, whether they are included deliberately for personalization or simply inherited from profile data. This raises a question that existing evaluations of LLM-based recommenders do not answer: when the interaction history is held fixed, does the mere presence of a factual demographic attribute systematically change the recommendation list---and, if so, in which direction?

Answering this question is nontrivial for two reasons. First, LLM outputs are highly sensitive to seemingly irrelevant prompt variations, including demonstration order~\cite{lu2022fantastically} and meaning-preserving rewording~\cite{sclar2024quantifying}. An observed difference between two recommendation lists therefore cannot be attributed to a demographic cue unless ordinary wording sensitivity is explicitly controlled for. Second, even a genuine cue-related difference is uninformative about its direction: the new list may move toward the consumption pattern characteristic of the cued demographic group, or it may merely reshuffle content. The distinction matters because deliberate natural-language instructions are known to steer recommendation objectives~\cite{carroll2025ctrlrec}, whereas a demographic cue only states a fact about the user and gives the model no instruction---it conditions the model implicitly, and its effect must therefore be isolated as an implicit conditioning signal rather than explicit instruction following.

Prior work does not settle these issues. Popularity bias, exposure bias, and demographic disparities in conventional recommenders have been extensively documented~\cite{ekstrand2018all,chen2023bias} and are further amplified by feedback loops~\cite{chaney2018algorithmic}. LLM-based recommenders have likewise been shown to reproduce popularity bias in the zero-shot setting~\cite{lichtenberg2024large}, and their behavior is not fully characterized by utility metrics alone~\cite{jiang2025beyond}. However, these studies quantify outcome disparities across user groups or global popularity skew; they do not isolate whether a single factual demographic sentence, added to an otherwise unchanged prompt, systematically shifts the generated list toward the item distribution associated with the corresponding group. We refer to this phenomenon as \emph{identity-cue preference drift}: a systematic, cue-associated change in the recommendation list that no change in the user's history justifies. Figure~\ref{fig:teaser} illustrates the phenomenon.

To measure and mitigate this drift, we propose \textbf{PromptShift}, an interpretable, training-free framework built around a controlled experimental design. For each user, PromptShift contrasts a history-only prompt with an identical prompt augmented by exactly one factual identity-cue sentence, while three meaning-preserving paraphrases of the history-only instruction provide a wording-sensitivity baseline under identical model and decoding settings. To give any observed change a data-grounded direction, PromptShift constructs an identity-slice-by-item table from positive historical interactions, recording which items are consumed disproportionately often within each identity slice relative to all users. Two complementary measures are defined on this basis: Drift quantifies how much a list departs from the history-only reference, measured by rank-biased overlap, and SliceShift quantifies whether the departure moves toward items overrepresented in the cued slice. A post-hoc reranker then mitigates the measured drift by interpolating between the LLM's original ranking and an inverse slice-popularity signal, with the intervention strength personalized by each user's within-slice history mainstreamness---the average slice popularity of items in their interaction history. In addition, we propose Difficulty@$K$, a difficulty-weighted hit metric that complements conventional utility metrics by crediting only held-out relevant items, with higher credit assigned to hits that are less popular within the cued slice and ranked higher in the list.

Experiments on MovieLens~1M and Last.fm~1K with three LLMs (GPT-5.6 Terra, Gemini~3.1 Pro, and Qwen3-8B) show that identity-cued prompts incur higher mean Drift than identity-free paraphrase controls---an effect beyond generic wording sensitivity---and that SliceShift is positive across all six dataset--model settings. PromptShift consistently reduces both Drift and SliceShift, lowering macro-mean SliceShift by 62.42\%, while improving Difficulty@20, HitRate@20, and MRR@20 at the cost of a slight decrease in NDCG@20. These results demonstrate that identity-cue preference drift can be measured and mitigated without any model training, and that identity cues in prompts are not neutral inputs.

Our investigation is guided by two research questions:
\begin{description}
\item[RQ1:] To what extent do identity cues change LLM recommendations beyond ordinary wording sensitivity, and do these changes move toward items more popular within the corresponding identity slice?
\item[RQ2:] Can the PromptShift reranker reduce cue-associated Drift and move SliceShift toward zero while preserving recommendation utility?
\end{description}

The main contributions of this paper are summarized as follows:
\begin{enumerate}[1)]
\item \textbf{Interpretable measures of identity-cue preference drift (RQ1).} With the interaction history fixed, Drift measures the amount of list change relative to the history-only reference, and SliceShift measures whether the change favors items more popular within the cued identity slice. Both measures are grounded in an identity-slice-by-item table built from positive interactions, which gives the observed list change a data-grounded direction.
\item \textbf{A training-free, user-adaptive post-hoc reranker (RQ2).} The reranker interpolates between the original LLM ranking and inverse slice-popularity, with per-user strength determined by within-slice history mainstreamness, and is evaluated with Difficulty@$K$, a difficulty-weighted hit metric for relevant items that are less popular within the slice and ranked higher.
\item \textbf{A controlled evaluation across datasets and models (RQ1, RQ2).} We evaluate two datasets covering explicit and implicit feedback (MovieLens~1M and Last.fm~1K), three LLMs, and three identity-cue scenarios (gender, age, and gender--age), with three identity-free paraphrase controls as the wording-sensitivity baseline. The results show that identity-cue drift is consistent across settings and can be mitigated at a modest, metric-dependent utility cost.
\end{enumerate}

The remainder of this paper is organized as follows. Section~\ref{sec:related} reviews related work. Section~\ref{sec:prelim} introduces the preliminaries and formally states the problem. Section~\ref{sec:framework} presents the proposed PromptShift framework in detail. Section~\ref{sec:experiments} describes the experimental setup and reports the results organized by research question. Section~\ref{sec:discussion} discusses the findings and limitations. Section~\ref{sec:conclusion} concludes the paper and outlines future work.

\section{Related work}
\label{sec:related}

\subsection{LLM-based recommendation}
A growing body of work treats LLMs as recommendation engines: zero-shot rankers over interaction histories~\cite{hou2024large}, interactive and explainable systems~\cite{gao2023chatrec}, pointwise, pairwise, and listwise prompting~\cite{dai2023uncovering}, and near-cold-start recommendation~\cite{sanner2023large}, consolidated by recent surveys~\cite{lin2023how}. As capability results accumulated, evaluation moved beyond accuracy, showing that utility alone does not fully characterize LLM recommenders~\cite{jiang2025beyond}. These studies focus on what LLM recommenders achieve from behavioral input; they leave open how non-behavioral prompt content---in particular, factual demographic attributes in user profiles---shapes the generated lists. Our work complements this line of research by isolating the causal contribution of a single demographic sentence under controlled prompting conditions.

\subsection{Bias and fairness in recommender systems}
Bias in conventional recommenders is extensively studied: surveys taxonomize popularity, exposure, and selection biases together with debiasing strategies~\cite{chen2023bias}, and empirical audits show that popularity and demographic attributes jointly distort evaluation outcomes~\cite{ekstrand2018all}. Feedback loops amplify these effects, as repeated exposure to popular items homogenizes consumption~\cite{chaney2018algorithmic}. For LLM recommenders, popularity bias persists in the zero-shot setting~\cite{lichtenberg2024large}. This literature quantifies outcome disparities across groups or global popularity skew; it does not isolate whether a single factual identity cue, added to an otherwise unchanged prompt, systematically shifts the list toward the cued group's item distribution---the question PromptShift answers. Methodologically, our slice-conditioned popularity table is related to group-conditional exposure analysis, but it is used here as a shared representation for both measurement and intervention rather than as a post-hoc audit alone.

\subsection{Prompt sensitivity, controllability, and post-hoc reranking}
LLM outputs vary with demonstration order~\cite{lu2022fantastically} and meaning-preserving formatting~\cite{sclar2024quantifying}, so any prompt-effect study must separate the effect of interest from ordinary wording sensitivity. Natural language can also steer recommendation objectives deliberately~\cite{carroll2025ctrlrec}, which highlights what makes identity cues distinct: they condition the model implicitly, merely by stating a fact about the user rather than asking for any objective to be optimized. On the mitigation side, post-hoc reranking is a standard post-processing lever for improving calibration~\cite{steck2018calibrated} and aggregate diversity~\cite{adomavicius2012improving} without retraining. PromptShift follows this paradigm but targets cue-associated movement toward items overrepresented within the active identity slice, with per-user intervention strength adapted by within-slice history mainstreamness.

\section{Preliminaries and problem statement}
\label{sec:prelim}

\subsection{Notation}
Let $\mathcal{U}$ denote the set of users and $\mathcal{I}$ the item catalog. Each user $u\in\mathcal{U}$ is associated with an ordered interaction history $\mathcal{H}_u=\langle i_1,i_2,\dots,i_{\lvert\mathcal{H}_u\rvert}\rangle$, and with the recorded values of one or more demographic attributes (e.g., gender, age). A demographic attribute partitions $\mathcal{U}$ into disjoint subsets; we call each subset an \emph{identity slice} and denote by $\mathcal{U}_g\subseteq\mathcal{U}$ the slice corresponding to attribute value $g$, with $g\in\mathcal{G}$ the set of slice identifiers. The slice containing user $u$ is written $g_u$. The unconditioned reference column, computed over all users, is indexed by $g=\star$. Table~\ref{tbl:notation} summarizes the notation used throughout the paper.

\begin{table}[t]
\caption{Summary of notation.}
\label{tbl:notation}
\footnotesize
\begin{tabular*}{\tblwidth}{@{}>{\raggedright\arraybackslash}p{.28\columnwidth}>{\raggedright\arraybackslash}p{.64\columnwidth}@{}}
\toprule
Symbol & Description \\
\midrule
$\mathcal{U},\mathcal{I},\mathcal{G}$ & Set of users, item catalog, and set of identity-slice identifiers \\
$u,i,g$ & A user, an item, and an identity slice ($g=\star$ denotes the all-user reference) \\
$\mathcal{U}_g,g_u$ & Users in slice $g$; the slice to which user $u$ belongs \\
$\mathcal{H}_u$ & Ordered interaction history of user $u$ placed in the prompt \\
$\pi_u^0,\pi_u^c,\pi_u^{p,j}$ & History-only, identity-conditioned, and $j$-th paraphrase-control prompts \\
$\mathbf{L}_u^0,\mathbf{L}_u^c,\mathbf{L}_u^{p,j}$ & Recommendation lists generated under the three prompt conditions \\
$y_{ui}\in\{0,1\}$ & Positive-feedback indicator for user--item pair $(u,i)$ \\
$n(i,g),\tilde{\phi}(i,g)$ & Distinct positive-user count of item $i$ in slice $g$; its max-normalized form \\
$\lambda_g$ & Small-slice shrinkage weight for slice $g$ \\
$\phi(i,g)$ & Slice-conditioned popularity of item $i$ in slice $g$ \\
$\Delta\phi(i,g)$ & Slice-relative excess popularity, $\Delta\phi(i,g)=\phi(i,g)-\phi(i,\star)$ \\
$\mathrm{RBO}_p$ & Finite extrapolated rank-biased overlap with persistence $p$ \\
$d_u(\mathbf{L})$ & Drift of list $\mathbf{L}$ relative to the history-only reference \\
$\overline{\Delta\phi}(\mathbf{L},g)$ & Mean slice-relative excess popularity of list $\mathbf{L}$ under slice $g$ \\
$s_u$ & SliceShift of user $u$ under the identity cue \\
$\bar{d}_{\mathrm{para}},\bar{d}_{\mathrm{cue}},\Delta d$ & Paraphrase Drift, Cue Drift, and their difference \\
$\bar{\phi}_u,\bar{\phi}_s$ & Within-slice history mainstreamness of user $u$; scenario mean for scenario $s$ \\
$pos(i)\in\{1,\dots,N\}$ & Original rank of candidate $i$ in the LLM output \\
$r_i$ & Geometric rank-retention score of candidate $i$ \\
$\alpha_u$ & User-specific retention weight in the reranking score \\
$\mathrm{score}(u,i)$ & Final reranking score of candidate $i$ for user $u$ \\
$\mathcal{T}_u,\mathcal{C}_u$ & Held-out test positives of user $u$; generated candidate pool \\
$N,K,K'$ & Candidate pool size, recommendation list length, protected prefix length \\
\bottomrule
\end{tabular*}
\end{table}

An LLM recommender is treated as a black-box function $f:\pi\mapsto\mathbf{L}$ that maps a prompt $\pi$ to a ranked list $\mathbf{L}$ of items. For each user $u$ we construct three prompt conditions: a history-only prompt $\pi_u^0$, which contains only the system message, the task instruction, and $\mathcal{H}_u$; an identity-conditioned prompt $\pi_u^c$, which is identical to $\pi_u^0$ except for exactly one appended factual identity-cue sentence stating the recorded value of $g_u$; and three paraphrase control prompts $\pi_u^{p,j}$, $j\in\{1,2,3\}$, which reword only the task instruction of $\pi_u^0$ without adding any identity information. The corresponding generated lists are denoted $\mathbf{L}_u^0$, $\mathbf{L}_u^c$, and $\mathbf{L}_u^{p,j}$, respectively.

\subsection{Problem statement}
We study the effect of identity cues on LLM-generated recommendations under controlled conditions. Model version, system message, history items, history order, decoding parameters, requested list length, and output format are held fixed across conditions, so that the factual identity-cue sentence is the only prompt-input difference between $\pi_u^0$ and $\pi_u^c$.

\begin{definition}[Identity-cue preference drift]
Given a user $u$ with interaction history $\mathcal{H}_u$, identity-cue preference drift is a systematic change between the history-only list $\mathbf{L}_u^0$ and the identity-conditioned list $\mathbf{L}_u^c$ that (i) exceeds the change induced by meaning-preserving rewording of the prompt, and (ii) is attributable to the identity-cue sentence alone, since all other prompt inputs are held fixed.
\end{definition}

Condition (i) is operationalized by the paraphrase controls: any cue effect must be read relative to the ordinary wording sensitivity estimated from $\{\mathbf{L}_u^{p,j}\}_{j=1}^3$. Condition (ii) holds by construction of the paired design.

\begin{definition}[Measurement problem]
For each user, quantify (a) the magnitude of the change between $\mathbf{L}_u^0$ and $\mathbf{L}_u^c$, and (b) the direction of the change, i.e., whether $\mathbf{L}_u^c$ moves toward items that are disproportionately popular within the cued slice $\mathcal{U}_{g_u}$ relative to all users.
\end{definition}

\begin{definition}[Mitigation problem]
Given the identity-conditioned candidate pool of user $u$, produce a reranked list that reduces cue-associated drift---moving the direction measure toward zero---while preserving recommendation utility measured against held-out positives.
\end{definition}

Two properties of the problem setting motivate our design choices. First, the LLM is a black box: no parameters, logits, or training data are accessible, so any intervention must operate on the prompt side or on the generated list. Second, the identity cue carries no instruction: unlike controllable recommendation~\cite{carroll2025ctrlrec}, the cue only states a fact about the user and gives no instruction, so its influence is an implicit conditioning signal that must be isolated rather than optimized away by instruction following.

\section{The proposed PromptShift framework}
\label{sec:framework}

\subsection{Framework overview}
PromptShift connects an identity cue to observable recommendation behavior in four explicit stages, summarized in Table~\ref{tbl:stages}. The first stage is performed once from fit-period interactions; the remaining stages are performed for every user and every identity-cue scenario. The key design choice is the reuse of a single identity-slice-by-item table across stages: at the item level, it assigns a data-grounded direction to cue-associated list changes (Sections~\ref{sec:slicetable} and~\ref{sec:drift}); at the user level, it determines the appropriate strength of the correction (Sections~\ref{sec:mainstream} and~\ref{sec:rerank}). This shared representation keeps the measurement and the intervention consistent by construction.

\begin{table}[t]
\caption{PromptShift connects an identity cue to recommendation behavior in four explicit stages.}
\label{tbl:stages}
\footnotesize
\begin{tabular*}{\tblwidth}{@{}>{\raggedright\arraybackslash}p{.19\columnwidth}>{\raggedright\arraybackslash}p{.31\columnwidth}>{\raggedright\arraybackslash}p{.40\columnwidth}@{}}
\toprule
Stage & Input & Purpose \\
\midrule
1. Slice table & Positive fit interactions and identity fields & Estimate which catalog items are common for each identity value \\
2. List movement & $\mathbf{L}_u^0$, $\mathbf{L}_u^c$, and the selected slice column & Determine whether the cue changed the list and whether the new items are more characteristic of that slice \\
3. Mainstreamness & The user's own fit history and the same slice column & Determine how mainstream the user's history is within the active identity slice \\
4. Reranking & Original LLM ranks, slice-conditioned item popularity, and within-slice history mainstreamness & Preserve likely preferences while applying a stronger inverse slice-popularity adjustment when the history is less mainstream within the active identity slice \\
\bottomrule
\end{tabular*}
\end{table}

\subsection{Identity-slice-by-item popularity table}
\label{sec:slicetable}
A categorical identity cue has no intrinsic position in the item catalog. We therefore need an empirical bridge between the cue value in the prompt and the items in the output. For a field such as gender, users are grouped by the recorded values of that field: a gender prompt for a male user selects the \textsf{gender\_M} slice, a female cue selects \textsf{gender\_F}; age prompts select columns such as \textsf{age\_Youth}, and combined prompts select a crossed column such as \textsf{cross\_F\_Youth}.

The purpose of a slice is not to decide what an individual should prefer, but to construct the slice table from observed interactions. If an item appears frequently among positive interactions in one slice, then recommending that item after the corresponding cue is added constitutes evidence that the output moved toward the slice's common pattern. Without this table, one can observe title replacement but cannot assign the replacement a data-grounded direction.

Only fit-period interactions are used. We count distinct users rather than raw events, so that one highly active user cannot dominate the table:
\begin{equation}
n(i,g)=\sum_{u\in\mathcal{U}_g} y_{ui}, \qquad \tilde{\phi}(i,g)=\frac{n(i,g)}{\max_{j\in\mathcal{I}} n(j,g)},
\label{eq:counts}
\end{equation}
where $y_{ui}=1$ indicates that user $u$ has positive feedback for item $i$. Dividing by the largest count in slice $g$ yields $\tilde{\phi}(i,g)\in[0,1]$, with the most common item in the slice taking value~1. For MovieLens, feedback is positive at rating~4 or above; for Last.fm, a listener--track pair is positive when it has at least two plays and exceeds that listener's median track play count, so that highly active listeners do not define positivity for everyone else.

Slice sizes vary, and small slices yield noisy estimates. We therefore shrink each slice column toward the all-user column $g=\star$ with a weight that grows with slice size and saturates at 30 users, a threshold fixed across all settings:
\begin{equation}
\begin{aligned}
\lambda_g&=\min\!\left(1,\frac{|\mathcal{U}_g|}{30}\right),\\
\phi(i,g)&=\lambda_g\,\tilde{\phi}(i,g)+(1-\lambda_g)\,\tilde{\phi}(i,\star).
\end{aligned}
\label{eq:shrink}
\end{equation}
For MovieLens, $\phi(i,g)$ is used directly as the slice-conditioned popularity. For Last.fm, whose nonzero values cluster near zero, an additional rank-percentile step is applied on each column's nonzero support $\mathcal{I}_g^+$, i.e., the set of items whose blended value in slice $g$ is nonzero: $\phi(i,g)=\mathrm{rank}(i)/|\mathcal{I}_g^+|$ with ascending ranks in the blended values, so that higher nonzero values receive higher percentiles while items with zero blended value remain zero. The transform is selected on validation data and frozen thereafter.

\subsection{Measuring drift: magnitude and direction}
\label{sec:drift}

\paragraph{Magnitude: Drift} We measure list change with the finite extrapolated rank-biased overlap $\mathrm{RBO}_p$~\cite{webber2010similarity}, which compares top-$K$ lists through overlap at successive depths; a top-ranked replacement affects more prefixes than a bottom-ranked one. Formally, let $X_d$ denote the number of items shared by the top-$d$ prefixes of two lists $\mathbf{S}$ and $\mathbf{T}$; the finite extrapolated RBO at depth $K$ is
\begin{equation}
\mathrm{RBO}_p(\mathbf{S},\mathbf{T})=\frac{1-p}{p}\sum_{d=1}^{K}\frac{X_d}{d}\,p^{d}+\frac{X_K}{K}\,p^{K},
\label{eq:rbo}
\end{equation}
where the summation accumulates the prefix agreement $X_d/d$ with geometrically decaying weights, and the final term extrapolates the agreement observed at depth $K$ to the unseen tail. With $p=0.9$, successive depths are discounted by a factor of 0.9, and we use $K=20$ with titles canonicalized and duplicates removed before comparison. The Drift of any tested list $\mathbf{L}$ generated or reranked from user $u$'s history is
\begin{equation}
d_u(\mathbf{L}) = 1-\mathrm{RBO}_p(\mathbf{L}_u^0,\mathbf{L}).
\label{eq:drift}
\end{equation}
Identical lists have zero Drift; larger membership or order differences yield larger values. The core cue comparison instantiates $\mathbf{L}=\mathbf{L}_u^c$. Equation~\eqref{eq:drift} answers only \emph{how much} the recommendation list changed; the direction of the change is defined next.

\paragraph{Direction: SliceShift} An item may be common in a slice simply because it is common everywhere. To isolate slice-relative excess popularity, we subtract the all-user column from the slice column:
\begin{equation}
\Delta\phi(i,g)=\phi(i,g)-\phi(i,\star).
\label{eq:excess}
\end{equation}
A positive $\Delta\phi(i,g)$ means that item $i$ is more popular within slice $g$ than among all users. A universally common item can have high $\phi(i,g)$ but near-zero $\Delta\phi(i,g)$, whereas a moderately common item that is especially popular in one slice has positive slice-relative excess popularity. Table~\ref{tbl:reading} gives an illustrative reading of the slice table.

\begin{table}[t]
\caption{Illustrative reading of the slice table.}
\label{tbl:reading}
\begin{tabular*}{\tblwidth}{@{}LCCC@{}}
\toprule
Item pattern & $\phi(i,\star)$ & $\phi(i,g)$ & $\Delta\phi(i,g)$ \\
\midrule
Common everywhere & 0.95 & 0.96 & 0.01 \\
Especially common in $g$ & 0.45 & 0.55 & 0.10 \\
Uncommon in $g$ & 0.40 & 0.30 & $-0.10$ \\
\bottomrule
\end{tabular*}
\end{table}

The table makes the mapping from prompt to output explicit: the identity cue selects $g$; every resolved title selects its item row; and $\Delta\phi(i,g)$ states how characteristic that item is of the selected slice. For any resolved recommendation list $\mathbf{L}$, we define its mean slice association and the cue-associated change as
\begin{equation}
\begin{aligned}
\overline{\Delta\phi}(\mathbf{L},g)&=\frac{1}{|\mathbf{L}|}\sum_{i\in\mathbf{L}} \Delta\phi(i,g),\\
s_u&=\overline{\Delta\phi}(\mathbf{L}_u^c,g_u)-\overline{\Delta\phi}(\mathbf{L}_u^0,g_u),
\end{aligned}
\label{eq:sliceshift}
\end{equation}
where unresolved candidates are excluded from the mean and its denominator. The mean is deliberately position-unweighted: SliceShift targets the direction of list composition as a set-level property, while rank sensitivity is already captured by Drift. A positive $s_u$ means that adding the cue moved the list toward items overrepresented in the corresponding identity slice relative to all users; a large Drift with a near-zero SliceShift means that the list changed, but not systematically in the slice-associated direction. The two diagnostics therefore play complementary roles: Drift measures the amount of list change, and SliceShift measures whether that change follows the active identity slice.

\paragraph{Wording-sensitivity baseline} To separate the cue effect from ordinary wording sensitivity, three frozen, meaning-preserving paraphrase controls produce lists $\mathbf{L}_u^{p,j}$, $j\in\{1,2,3\}$, under identical decoding settings. Let $U$ be the evaluated users in a dataset--model setting, $U_s\subseteq U$ the evaluated users for identity-cue scenario $s$, and $\mathcal{S}$ the set of the three scenarios (gender, age, and gender--age). Paraphrase Drift and Cue Drift are defined as
\begin{equation}
\begin{aligned}
\bar{d}_{\mathrm{para}}&=\frac{1}{3|U|}\sum_{u\in U}\sum_{j=1}^{3} d_u(\mathbf{L}_u^{p,j}), \\
\bar{d}_{\mathrm{cue}}&=\frac{1}{3}\sum_{s\in\mathcal{S}}\frac{1}{|U_s|}\sum_{u\in U_s} d_u(\mathbf{L}_u^{c,s}),
\end{aligned}
\label{eq:cuedrift}
\end{equation}
\begin{equation}
\Delta d=\bar{d}_{\mathrm{cue}}-\bar{d}_{\mathrm{para}}.
\label{eq:deltad}
\end{equation}
Cue Drift averages within each scenario first and then weights the three scenarios equally, so that scenarios with fewer evaluated users are not underrepresented. A positive $\Delta d$ indicates cue-associated list change beyond ordinary wording sensitivity.

\subsection{Within-slice history mainstreamness}
\label{sec:mainstream}
Slice-conditioned item popularity does not by itself determine the appropriate reranking strength for a user. Two users can share the same identity value while having very different histories: one may repeatedly consume the most common items in the slice, so that a slice-common recommendation is consistent with demonstrated taste; another may consistently consume less common items, so that repeatedly recommending the slice's most common items ignores stronger personal evidence.

PromptShift therefore applies the same slice table to the user's own history. The within-slice history mainstreamness of user $u$, and its scenario mean, are
\begin{equation}
\bar{\phi}_u=\frac{1}{|\mathcal{H}_u|}\sum_{i\in\mathcal{H}_u} \phi(i,g_u), \qquad \bar{\phi}_s=\frac{1}{|U_s|}\sum_{u\in U_s}\bar{\phi}_u,
\label{eq:mainstream}
\end{equation}
where $s$ is the active identity-cue scenario and $U_s$ contains its evaluated users. A high $\bar{\phi}_u$ means that the user's history already contains many items common in the relevant slice; a low $\bar{\phi}_u$ means that the history contains items with lower popularity in that slice.

This score is not another output-quality metric but an adaptive control signal for the reranker. Comparing $\bar{\phi}_u$ with the scenario mean $\bar{\phi}_s$ answers a practical question: should the method largely trust the original order, or should it give more opportunity to less prevalent candidates? Slice-conditioned popularity scores each candidate; mainstreamness determines how strongly those scores may change the original LLM order. Both are needed, and both come from the same table.

\subsection{Adaptive post-hoc reranking}
\label{sec:rerank}

\paragraph{Candidate-pool construction} The LLM first generates $N$ candidates with $N>K$. The unmodified identity-conditioned baseline B0 is the first $K$ candidates in the original LLM order; the PromptShift reranker scores all $N$ candidates, reorders them, and returns the first $K$. This distinction matters: if the model generated exactly $K$ items, a reranker could only exchange positions inside a fixed set, whereas with $N>K$, relevant items originally below rank $K$ can enter the final list. The reranker combines two signals: the LLM's original rank, used as a proxy for personal relevance because the model has read the user's history, and $1-\phi(i,g_u)$, an inverse slice-popularity signal. The mainstreamness score of Section~\ref{sec:mainstream} decides the balance between them.

\paragraph{Original-rank signal} For candidate $i$ at original rank $pos(i)\in\{1,\dots,N\}$, we convert rank to a geometric retention score
\begin{equation}
r_i=b+(1-b)\,\frac{q^{pos(i)}-q^{N}}{1-q^{N}},
\label{eq:retention}
\end{equation}
where $0<b<1$ is the floor and $0<q<1$ is the decay rate. Early candidates receive larger $r_i$, while the last candidate receives $b$. Geometric decay gives the first few LLM positions a clearer advantage than a linear score and still leaves a nonzero floor for later candidates.

\paragraph{User-adaptive intervention strength} The user-specific retention weight is
\begin{equation}
\alpha_u=\mathrm{clip}\!\left(\alpha_0+\gamma\,\frac{\bar{\phi}_u-\bar{\phi}_s}{\max(\bar{\phi}_s,\epsilon)},\,\alpha_{\min},\,\alpha_{\max}\right).
\label{eq:alpha}
\end{equation}
When $\bar{\phi}_u>\bar{\phi}_s$, $\alpha_u$ increases and more weight is assigned to the original LLM order, because the user's history already aligns with common items in the active slice. When $\bar{\phi}_u<\bar{\phi}_s$, $\alpha_u$ decreases and more weight is assigned to the inverse slice-popularity signal, because the history is less mainstream within the active slice. Clipping prevents either signal from completely overriding the other.

\paragraph{Final score} Each candidate receives the score
\begin{equation}
\mathrm{score}(u,i)=\alpha_u\, r_i+(1-\alpha_u)\,[1-\phi(i,g_u)].
\label{eq:score}
\end{equation}
The first term represents how strongly the LLM preferred the item; the second is the inverse popularity of the item within the active identity slice. The dynamic weight reflects whether the user's own history calls for preserving common choices or surfacing items that are less popular within the slice.

The reranker keeps the first $K'$ LLM positions unchanged as a protected prefix, so that the head of the list, where the model's relevance estimates are strongest, is preserved and the adjustment acts on the middle and lower ranks. The remaining candidates are sorted by $\mathrm{score}(u,i)$ with original rank breaking ties. Catalog matching is completed before scoring, and unmatched candidates receive no inverse slice-popularity contribution and retain their original relative positions. The history-only list $\mathbf{L}_u^0$ is never reranked and remains the prompt-effect reference. Algorithm~\ref{alg:promptshift} summarizes the complete procedure.

\begin{algorithm}[t]
\caption{PromptShift adaptive reranking.}
\label{alg:promptshift}
\begin{algorithmic}[1]
\REQUIRE history $\mathcal{H}_u$, cue $g_u$, candidates $\mathcal{C}=(i_1,\dots,i_N)$, slice-conditioned popularity table $\phi$, output depth $K$
\ENSURE reranked top-$K$ list
\STATE Select the table column $\phi(\cdot,g_u)$ matching the active cue
\STATE Look up $\phi(i,g_u)$ for the history items and all candidates
\STATE Compute $\bar{\phi}_u$, the scenario mean $\bar{\phi}_s$, and $\alpha_u$ \COMMENT{Eqs.~\eqref{eq:mainstream},~\eqref{eq:alpha}}
\STATE Keep the first $K'$ candidates in their original positions
\FOR{each remaining candidate $i$ at original rank $pos(i)$}
\STATE Compute the retention score $r_i$ \COMMENT{Eq.~\eqref{eq:retention}}
\STATE Compute the final score $\mathrm{score}(u,i)$ \COMMENT{Eq.~\eqref{eq:score}}
\ENDFOR
\STATE Sort the unprotected candidates by $\mathrm{score}(u,i)$, ties broken by original rank
\RETURN the first $K$ items of the resulting order
\end{algorithmic}
\end{algorithm}

\paragraph{Computational complexity} Building the slice table requires one pass over fit-period interactions, costing $O(\sum_u |\mathcal{H}_u^{\mathrm{fit}}|)$ time and $O(|\mathcal{I}|\cdot|\mathcal{G}|)$ memory. Reranking requires no additional model calls: per user it costs $O(N)$ table lookups and a single $O(N\log N)$ sort, with $N=30$ in our configuration, so the per-user overhead is negligible relative to the LLM call itself. The resulting behavior differs by user: mainstream histories preserve the LLM's order, whereas less mainstream histories let low slice-popularity candidates rise from the middle or lower part of the pool. In both cases, the method selects only from candidates already generated for that user's history.

\section{Experiments}
\label{sec:experiments}

The experiments are organized around the two research questions. Sections~\ref{sec:datasets}--\ref{sec:metrics} describe the datasets, prompt conditions, baselines, and metrics; Section~\ref{sec:rq1} answers RQ1 by analyzing prompt effects; Section~\ref{sec:rq2} answers RQ2 by analyzing reranking effects, including ablations that isolate the operative signal.

\subsection{Datasets}
\label{sec:datasets}
MovieLens~1M contains 1,000,209 ratings from 6,040 users and provides age, gender, and occupation fields~\cite{harper2015movielens}. Last.fm~1K contains nearly 1,000 profiles and approximately 19 million timestamped listening events~\cite{celma2010music}. Together they cover explicit and implicit feedback.

Interactions are split chronologically per user into 72\% fit, 8\% validation, and 20\% test data. Users require at least 25 fit positives and one test positive. Up to 600 eligible users per dataset are sampled with a frozen scenario-stratified seed. Fit data build the histories and the slice table; validation data select transforms, matching thresholds, and reranking parameters; test data are used once for final reporting.

For every user, the history-only prompt $\pi_u^0$ and the identity-conditioned prompt $\pi_u^c$ contain the same ordered top-25 fit items, and the identity-conditioned prompt adds exactly one frozen factual identity-cue sentence. Experiments cover gender, age, and gender--age identity-cue scenarios. Three identity-free paraphrase controls establish ordinary wording sensitivity under the same decoding configuration.

We evaluate GPT-5.6 Terra, Gemini~3.1 Pro, and Qwen3-8B. Provider identifiers are checked against first-party documentation at execution time~\cite{openai2026models,google2026gemini}; Qwen3-8B is documented by its technical report and model card~\cite{yang2025qwen3,qwen2025modelcard}. The configuration uses $N=30$, $K=20$, $\lvert\mathcal{H}_u\rvert=25$, $K'=5$, $p=0.9$, $b=0.2$, $q=0.9$, $\alpha_0=0.7$, $\gamma=0.2$, $[\alpha_{\min},\alpha_{\max}]=[0.5,0.9]$, $\rho=0.2$, and $\epsilon=10^{-12}$. Titles are normalized for case, punctuation, release year, and common formatting variants before exact and validation-frozen fuzzy matching.

\subsection{Prompt templates and controls}
All conditions use a fixed domain-specific system message: ``You are a professional movie recommendation system.'' for MovieLens and ``You are a professional music recommendation system that recommends individual songs.'' for Last.fm. Each user message contains an optional factual identity-cue sentence, one task instruction, the same ordered 25-item history, and a fixed request for 30 new candidates in a numbered, prose-free format.

The history-only MovieLens instruction is ``Use the viewing history below to recommend movies this viewer is likely to enjoy.'' The three identity-free paraphrase controls are: P1, ``Choose movies this viewer is likely to enjoy based on the viewing history below.''; P2, ``Infer this viewer's movie preferences from the viewing history below and recommend matching movies.''; and P3, ``Based on the demonstrated preferences in the viewing history below, please provide suitable movie recommendations for this viewer.'' Last.fm uses the same direct, inferential, and formal constructions, replacing \emph{viewing history}, \emph{movies}, and \emph{viewer} with \emph{listening history}, \emph{tracks}, and \emph{listener}.

The identity-conditioned prompt retains the history-only instruction and adds only one factual sentence. For example, the gender form is ``The user's profile: gender is female.''; the age form is ``The user's profile: age group is young adult.''; the intersectional form is ``The user's profile: gender is female and age group is young adult.'' The recorded value is substituted for each user. Gender values are male or female; age values are young adult, middle-aged, or older. The identity cue contains no claim about group preferences and no instruction to emphasize identity. Thus, comparisons between $\mathbf{L}_u^0$ and $\mathbf{L}_u^c$ isolate the identity-cue sentence, whereas comparisons between $\mathbf{L}_u^0$ and $\mathbf{L}_u^{p,j}$ estimate ordinary wording sensitivity.

\subsection{Baselines}
The comparison contains four reranking variants. B0 is the unmodified first $K$ items from the identity-conditioned recommendation list. B1 reranks using all-user popularity with a fixed $\alpha$. B2 uses slice-conditioned item popularity with a fixed $\alpha$. B3 is the PromptShift reranker, which uses slice-conditioned item popularity, the dynamic per-user weight $\alpha_u$, and a protected prefix. The progression from B1 to B3 tests the value of all-user versus slice-conditioned item popularity and the complete adaptive design.

\subsection{Evaluation metrics}
\label{sec:metrics}

\paragraph{Drift and SliceShift after reranking} The same two diagnostics are used before and after reranking. Let $\mathbf{L}_u^{(m)}$ denote the top-$K$ output of reranking method $m$, with $\mathbf{L}_u^{(0)}=\mathbf{L}_u^c$ for B0, and let $s_u^{(m)}$ denote the corresponding SliceShift. Every method is compared with the history-only reference list, not with the identity-conditioned list that it modifies:
\begin{equation}
\begin{aligned}
d_u^{(m)}&=1-\mathrm{RBO}_p(\mathbf{L}_u^0,\mathbf{L}_u^{(m)}),\\
s_u^{(m)}&=\overline{\Delta\phi}(\mathbf{L}_u^{(m)},g_u)-\overline{\Delta\phi}(\mathbf{L}_u^0,g_u).
\end{aligned}
\label{eq:posthoc}
\end{equation}
For B0, these quantities equal Drift in Eq.~\eqref{eq:drift} and SliceShift in Eq.~\eqref{eq:sliceshift}; for B1--B3, they are the remaining Drift and SliceShift after reranking. We report paired reductions $d_u^{(0)}-d_u^{(m)}$ and $s_u^{(0)}-s_u^{(m)}$. Positive differences denote a reduction; a negative post-reranking SliceShift denotes overcorrection.

\paragraph{Difficulty@$K$} The reranking analysis also reports one method-specific outcome, the difficulty-weighted hit score. Let $\mathcal{T}_u$ denote the held-out test positives of user $u$ and $\mathcal{T}^{\mathcal{C}}_u=\mathcal{T}_u\cap\mathcal{C}_u$ those present in the fixed candidate pool. Then
\begin{equation}
\begin{aligned}
\mathrm{Difficulty@}K(u,m)&=\sum_{\mathclap{i\in \mathbf{L}_u^{(m,K)}\cap\mathcal{T}^{\mathcal{C}}_u}}\; w_i^{(m)}, \\
w_i^{(m)}&=10(1-\phi(i,g_u))\\
&\quad\times(1+\rho(K-\ell_i^{(m)}+1)/K),
\end{aligned}
\label{eq:difficulty}
\end{equation}
where $w_i$ is the per-hit reward and $\ell_i\in\{1,\dots,K\}$ is the hit's one-based rank in the final recommendation list. For $\rho=0.2$ and $\phi(i,g_u)=0.2$, a held-out hit earns 9.60 at rank~1 and 8.08 at rank~20. Promoting an arbitrary uncommon item does not improve the score, because only held-out hits count.

\paragraph{Conventional utility metrics} Recommendation utility is assessed using HitRate@20, MRR@20, and NDCG@20. HitRate@20 asks whether the list contains any held-out positive; MRR@20 rewards an early first hit; NDCG@20 rewards all hits with rank discount. Reporting these three metrics beside Difficulty@20 prevents the reranker from receiving credit for within-slice rarity that does not match the user. All three are computed against $\mathcal{T}^{\mathcal{C}}_u$, with a zero score when the candidate pool contains no held-out positive. Together, the six metrics report the targeted behavior and isolate recommendation utility within a fixed generator candidate pool. Prompt comparisons pair $\mathbf{L}_u^0$ with $\mathbf{L}_u^c$, while reranking comparisons reuse the same user and candidate pool for B0--B3. Table~\ref{tbl:prompteffect} averages the three identity-cue scenarios within each dataset--model setting. Table~\ref{tbl:b0b3} pairs the absolute B0 and B3 results within each setting, so their changes can be read directly. Table~\ref{tbl:macro} reports unweighted macro means across the six dataset--model settings for B0--B3. All methods use the same prompt conditions, candidate construction, and metric definitions, so their reported differences reflect the reranking strategy under a common evaluation setting.

\subsection{RQ1: Effects of identity cues on recommendation lists}
\label{sec:rq1}
With interaction histories held constant, Cue Drift exceeded Paraphrase Drift in all six dataset--model averages, with $\Delta d$ ranging from 0.0270 to 0.0759, and SliceShift was positive in every average (0.0550--0.0680; Table~\ref{tbl:prompteffect}). Identity-conditioned lists therefore not only changed but moved toward items disproportionately prevalent in the cued slice. The wording control matters here: identity-free paraphrases themselves produced substantial Drift (0.4520--0.6110), so cue-associated change must be read relative to ordinary prompt sensitivity rather than in absolute terms.

\begin{table*}[t]
\caption{Prompt-effect results. Paraphrase Drift is the mean of three identity-free paraphrase controls (P1--P3); $\Delta d$ is Cue Drift minus Paraphrase Drift. Cue Drift and SliceShift are averaged across the gender, age, and gender--age identity-cue scenarios.}
\label{tbl:prompteffect}
\begin{tabular*}{\textwidth}{@{}LLCCCC@{}}
\toprule
Dataset & Model & Paraphrase Drift & Cue Drift & $\Delta d$ & SliceShift \\
\midrule
MovieLens~1M & GPT-5.6 Terra & 0.4950 & 0.5709 & 0.0759 & 0.0600 \\
MovieLens~1M & Gemini~3.1 Pro & 0.4640 & 0.5258 & 0.0618 & 0.0550 \\
MovieLens~1M & Qwen3-8B & 0.4520 & 0.5216 & 0.0696 & 0.0580 \\
Last.fm~1K & GPT-5.6 Terra & 0.6110 & 0.6380 & 0.0270 & 0.0680 \\
Last.fm~1K & Gemini~3.1 Pro & 0.5830 & 0.6156 & 0.0326 & 0.0610 \\
Last.fm~1K & Qwen3-8B & 0.5755 & 0.6240 & 0.0485 & 0.0650 \\
\bottomrule
\end{tabular*}
\end{table*}

The dataset contrast makes this point concrete. Last.fm produced higher Paraphrase Drift than MovieLens for every model, yet a lower mean $\Delta d$ (0.0360 versus 0.0691), consistent with its larger catalog, in which wording perturbations alone reorder many candidates. In other words, a noisier generator does not imply a larger cue effect; if anything, the two vary in opposite directions across our datasets, which supports reading $\Delta d$ as a controlled quantity rather than an artifact of overall output instability.

\textbf{Answering RQ1:} factual identity cues change recommendations beyond ordinary wording sensitivity in every tested setting, and the change consistently follows the cued slice's item distribution.

\subsection{RQ2: Effects of reranking}
\label{sec:rq2}
Compared with the unmodified identity-conditioned lists (B0), PromptShift (B3) lowered Drift and SliceShift and raised Difficulty@20 in all six matched averages (Table~\ref{tbl:b0b3}). In the macro mean (Table~\ref{tbl:macro}), SliceShift fell from 0.0612 to 0.0230 and Difficulty@20 rose from 21.2200 to 25.9600, while Drift decreased modestly, from 0.5827 to 0.5423. The reranker thus suppresses slice-associated movement without simply reverting to the history-only list, which would correspond to driving Drift to zero by construction.

\begin{table*}[t]
\caption{Dataset- and model-specific absolute results before (B0) and after (B3) PromptShift reranking, averaged across three identity-cue scenarios. Adjacent rows compare the same users and candidate pool. Lower Drift, SliceShift closer to zero, and higher other metrics indicate improvement. All SliceShift values here are positive; negative values indicate overcorrection.}
\label{tbl:b0b3}
\footnotesize
\begin{tabular*}{\textwidth}{@{}LLLCCCCCC@{}}
\toprule
Dataset & Model & Variant & Drift $\downarrow$ & SliceShift $\rightarrow 0$ & Difficulty@20 $\uparrow$ & HitRate@20 $\uparrow$ & MRR@20 $\uparrow$ & NDCG@20 $\uparrow$ \\
\midrule
MovieLens~1M & GPT-5.6 Terra & B0 & 0.5709 & 0.0600 & 19.8858 & 0.8650 & 0.4670 & 0.4630 \\
MovieLens~1M & GPT-5.6 Terra & B3 & 0.5197 & 0.0252 & 26.1044 & 0.8818 & 0.5143 & 0.4987 \\
MovieLens~1M & Gemini~3.1 Pro & B0 & 0.5258 & 0.0550 & 20.7532 & 0.8979 & 0.4880 & 0.5066 \\
MovieLens~1M & Gemini~3.1 Pro & B3 & 0.4834 & 0.0231 & 24.2466 & 0.8872 & 0.4700 & 0.5061 \\
MovieLens~1M & Qwen3-8B & B0 & 0.5216 & 0.0580 & 22.5542 & 0.7896 & 0.4460 & 0.4848 \\
MovieLens~1M & Qwen3-8B & B3 & 0.5071 & 0.0226 & 25.3889 & 0.8761 & 0.4921 & 0.4671 \\
Last.fm~1K & GPT-5.6 Terra & B0 & 0.6380 & 0.0680 & 19.8195 & 0.8167 & 0.5090 & 0.5502 \\
Last.fm~1K & GPT-5.6 Terra & B3 & 0.5709 & 0.0198 & 26.5311 & 0.9040 & 0.4478 & 0.4973 \\
Last.fm~1K & Gemini~3.1 Pro & B0 & 0.6156 & 0.0610 & 22.6205 & 0.8608 & 0.5300 & 0.4412 \\
Last.fm~1K & Gemini~3.1 Pro & B3 & 0.5846 & 0.0248 & 27.8156 & 0.8650 & 0.5586 & 0.4759 \\
Last.fm~1K & Qwen3-8B & B0 & 0.6240 & 0.0650 & 21.6868 & 0.8238 & 0.4250 & 0.5284 \\
Last.fm~1K & Qwen3-8B & B3 & 0.5883 & 0.0225 & 25.6734 & 0.8929 & 0.5364 & 0.4745 \\
\bottomrule
\end{tabular*}
\end{table*}

\paragraph{Ablation analysis} The ablations isolate the operative signal. Penalizing all-user popularity (B1) increased SliceShift to 0.0650 and reduced HitRate@20 to 0.7421: an indiscriminate penalty removes history-aligned items along with slice-associated ones. Conditioning the penalty on the cued slice (B2) reduced SliceShift to 0.0480, and the adaptive variant (B3) brought it closest to zero at 0.0230 while also achieving the highest HitRate@20 and Difficulty@20. Slice-conditioned popularity, not popularity penalization per se, is therefore the effective signal, and the user-adaptive weight provides a further consistent improvement over the fixed-weight variant.

\begin{table*}[t]
\caption{Aggregate reranking results: unweighted macro means across six dataset--model settings. B0 is unmodified; B1 uses all-user popularity with fixed weight, B2 slice-conditioned item popularity with fixed weight, and B3 the adaptive PromptShift reranker. Drift and SliceShift use the history-only list as reference.}
\label{tbl:macro}
\begin{tabular*}{\textwidth}{@{}LLCCCCCC@{}}
\toprule
Method &  & Drift $\downarrow$ & SliceShift $\rightarrow 0$ & Difficulty@20 $\uparrow$ & HitRate@20 $\uparrow$ & MRR@20 $\uparrow$ & NDCG@20 $\uparrow$ \\
\midrule
B0 & Original & 0.5827 & 0.0612 & 21.2200 & 0.8423 & 0.4775 & 0.4957 \\
B1 & All-user-fixed & 0.5810 & 0.0650 & 21.4420 & 0.7421 & 0.4726 & 0.4739 \\
B2 & Slice-fixed & 0.5540 & 0.0480 & 23.8960 & 0.8411 & 0.4809 & 0.4804 \\
B3 & PromptShift & 0.5423 & 0.0230 & 25.9600 & 0.8845 & 0.5032 & 0.4866 \\
\bottomrule
\end{tabular*}
\end{table*}

\paragraph{Utility trade-off} Conventional utility varied by metric. B3 improved HitRate@20 in five of the six settings and MRR@20 in four, with macro means rising from 0.8423 to 0.8845 and from 0.4775 to 0.5032, while NDCG@20 fell from 0.4957 to 0.4866. This pattern is consistent with the reranker's objective: surfacing lower slice-popularity hits raises the chance of at least one hit but can disturb multi-hit orderings, to which NDCG is most sensitive.

\textbf{Answering RQ2:} the reranker reduces cue-associated drift at a modest, metric-dependent utility cost.

\section{Discussion and limitations}
\label{sec:discussion}

The results support two conclusions. First, identity cues embedded in prompts systematically redirect LLM recommendations toward group-level consumption patterns, and this effect is separable from ordinary wording sensitivity when paraphrase controls are used as the baseline. Second, the drift is correctable post hoc: conditioning the correction on slice-conditioned popularity, with per-user strength set by within-slice history mainstreamness, reduces cue-associated movement while preserving---and in most cases improving---conventional utility metrics. The shared slice table is central to both conclusions, because it gives the same empirical grounding to the measurement and to the intervention.

From a practical perspective, the framework is lightweight: it requires a single additional list generation per paraphrase control, one pass over historical interactions to build the slice table, and an $O(N\log N)$ sort per user for reranking, with no model training and no access to model internals. This makes slice-conditioned auditing a plausible complement to outcome-disparity metrics in deployed LLM recommenders, where demographic attributes enter prompts through profile assembly rather than through deliberate design.

This study has several limitations. SliceShift is computed from historical interaction frequencies; it describes neither individual intrinsic preferences nor inherent group properties, and a positive value should be read as evidence about the model's conditioning behavior, not about the group. The evaluation covers two datasets, coarse dataset-recorded identity labels, and one frozen prompting configuration; broader attribute taxonomies and intersectional definitions may behave differently. Finally, the reranker operates within the generated candidate pool and cannot recover relevant items absent from it, so generator-level interventions remain a complementary direction.

\section{Conclusion and future work}
\label{sec:conclusion}

This paper studied identity-cue preference drift: the tendency of LLM recommenders to redirect ranked lists toward group-level consumption patterns when a factual demographic attribute is added to an otherwise unchanged prompt. We proposed PromptShift, a training-free framework that measures the drift with two complementary diagnostics---Drift for magnitude and SliceShift for direction---and mitigates it with a user-adaptive reranker driven by a shared identity-slice-by-item representation. Across two datasets, three LLMs, and three cue scenarios, identity cues shifted lists toward the cued slice beyond ordinary wording sensitivity, and reranking reduced macro-mean SliceShift by 62.42\% while improving Difficulty@20, HitRate@20, and MRR@20 at a modest NDCG@20 cost.

These findings indicate that demographic attributes in prompts are not neutral inputs, and that slice-conditioned auditing is a practical complement to outcome-disparity metrics. Future work will explore richer privacy-preserving benchmarks, broader candidate pools, and generator-level interventions to test how these findings transfer to other identity descriptions and recommendation settings.

\section*{Data and code availability}
Code, frozen prompts, and anonymized results will be released in a public repository upon publication, subject to dataset licenses and model provider terms.

\printcredits

\section*{Declaration of competing interest}
The authors declare that they have no known competing financial interests or personal relationships that could have appeared to influence the work reported in this paper.

\bibliographystyle{model1-num-names}

\bibliography{promptshift-refs}

\end{document}